\documentclass[preprint,prl]{revtex4}

\usepackage{epsfig}

\begin{document}

\title{Ergodicity Breaking in a Deterministic Dynamical System}
\author{Golan Bel and Eli Barkai}

\affiliation{ Physics Department, Bar-Ilan University, Ramat-Gan 52900, Israel }

\begin{abstract}
The concept of weak ergodicity breaking is defined and studied in the context of deterministic dynamics.
We show that weak ergodicity breaking describes a weakly chaotic dynamical system: a nonlinear
map which generates subdiffusion deterministically.
In the non-ergodic phase non-trivial distribution of the fraction of occupation times is obtained.
The visitation fraction remains uniform even in the non-ergodic phase.
In this sense the non-ergodicity is quantified, leading to a statistical
mechanical description of the system even though it is not ergodic.
\end{abstract}

\maketitle

\section{Introduction}
The relation between deterministic non-linear dynamics and statistical
mechanics is a topic of wide interest \cite{Dorfman,Zas}. In
particular, the domain of validity of the ergodic hypothesis for low
dimensional dynamical systems is fundamental in our understanding of the
foundations of statistical mechanics. In usual setting of statistical
mechanics the phase space of a closed system is divided into equally sized
cells, then if the system is ergodic it will spend equal amount of time in
each cell in the limit of long measurement time. Ofcourse many dynamical
systems are non-ergodic, for example the phase space may decompose into
regions, where the system cannot access one part of the phase space if it
started in a different part. Such non-ergodicity is called here strong non-ergodicity.\newline
\indent A second type of non-ergodicity, called weak 
non-ergodicity is investigated in this manuscript. As conjectured by Bouchaud \cite{Bouchaud1} in the context
of glassy dynamics, in weak ergodicity breaking the phase space is \textit{a priori} not broken into mutually inaccessible regions.
However the system does not spend equal amount of time in equally sized regions of phase space (see detailed
definition below). While occupation times of cells in phase space remain random, we showed in the context
of \emph{stochastic} dynamics that distribution of occupation times is universal \cite{Bel} in a way that will be made clearer later.
The goal of this paper is to investigate weak non-ergodicity for a \emph{deterministic} process.\newline
\indent In terms of stochastic theories like the trap model, the random energy model \cite{Bouchaud1} and the continuous time random
walk \cite{Bel} it is known that when the dynamics exhibits aging and anomalous diffusion it may exhibit
weak ergodicity breaking. It is also well known that a
wide variety of deterministic dynamical systems exhibit anomalous diffusion
\cite{Zaslavski1,Klafter1,Tasaki,Arutso}. Thus the question arises: Can
deterministic dynamical systems exhibit weak ergodicity breaking?
In this manuscript a detailed definition of weak ergodicity breaking (WEB) in the context of deterministic dynamics 
is introduced. It is shown that that a low dimensional model, a deterministic non linear map
which has no disorder built into it, yields WEB. Non-trivial (in a way defined
later) probability density function (PDF) of fraction of occupation time is shown to describe equilibrium
properties of the dynamical system. In this
sense the non-ergodicity is quantified, enabling us to perform statistical mechanical description of
the system although it is not ergodic.
\section{Geisel's Maps} 
\begin{figure}[tbp]
\begin{center}
\epsfxsize=80mm \epsfbox{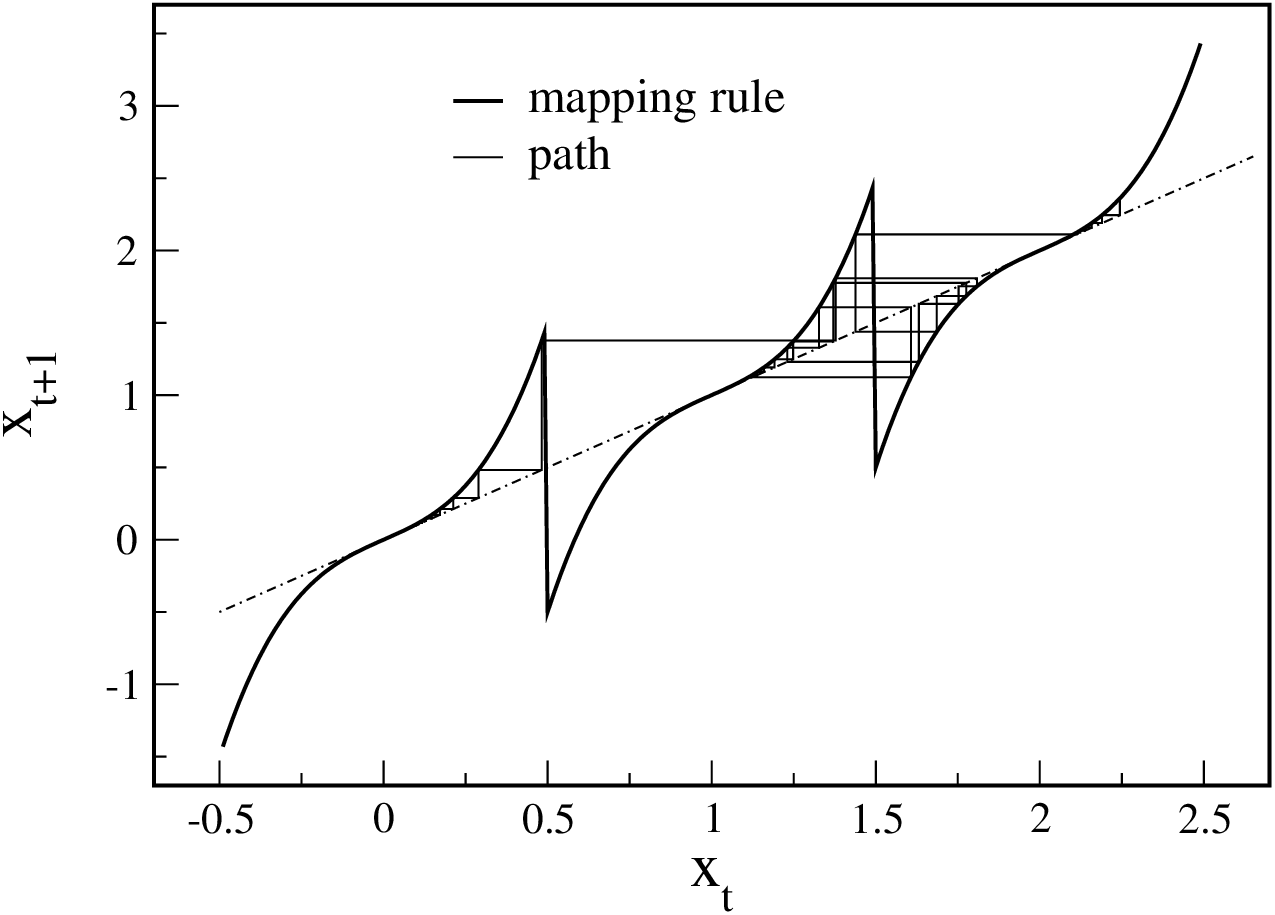}
\end{center}
\caption{ The mapping Eqs. (\ref{n1},\ref{n3}) with z=3 and L=3, the
dashed line shows the curve $x_{t+1}=x_{t} $. Fixed points are at $x_{t}=0,1,2$.
A specific realization of the dynamics
is also drawn (thin line). }
\label{fig1}
\end{figure}

Probably the simplest deterministic maps which lead to normal and
anomalous diffusion are one dimensional maps of the form%
\begin{equation}
x_{t+1}=x_{t}+F\left( x_{t}\right),   \label{n1}
\end{equation}%
with the following symmetry properties of $F\left( x\right) :$ (i) $F\left(
x\right) $ is periodic with a period interval set to $1$, i.e., $F\left(
x\right) =F\left( x+N\right) ,$ where N is an integer. (ii) $F\left(
x\right) $ is anti-symmetric; namely, $F\left( x\right) =-F\left( -x\right) ,
$ note that $t$ in Eq. (\ref{n1}) is a discrete time. Geisel and Thomae \cite{Geisel1}
considered a wide family of such maps which behave like%
\begin{equation}
F\left( x\right) =ax^{z}\mbox{\ \ for \ \ }x\rightarrow 0_{+}  \label{n2}
\end{equation}%
where $z>1$. Eq. (\ref{n2}) defines the property of the map close to its
unstable fixed point. In numerical experiments introduced later we will use the following map%
\begin{equation}
F\left( x\right) =\left( 2x\right) ^{z},\mbox{ \ \ \ \ }0\leq x\leq 1/2
\label{n3}
\end{equation}%
which together with the symmetry properties of the map define the mapping
for all $x$. The functional form of the map on a unit interval was introduced by
Pomeau and Manneville to describe intermittency \cite{PM}. Variations of these maps
have been investigated by taking into account time dependent noise
\cite{Bettin}, quenched disorder \cite{Radons}, and additional uniform bias which 
breaks the symmetry of the map \cite{Barkai2}.
We consider the map with $L $ unit cells and periodic boundary conditions. 
In Fig. \ref{fig1} we show the map, and one realization of a path along three unit cells.\newline
\indent The dynamics of the map in the vicinity of the unstable fixed points, is
governed by a \textit{q exponential} sensitivity of the trajectories to initial
conditions, i.e., weak chaos. The role of such \textit{q exponential} in models
related to ours (the logistic map) and Tsallis statistics is a subject of ongoing research \cite{Baldovin,Robledo,Ananos}.
Grassberger \cite{Grass} claimed that ideas related to Tsallis statistics are wrong due to
misconceptions on basic notions of ergodic theory. here we show that for the
weakly chaotic map under consideration, the correct statistical framework is
given by the concept of WEB.\newline
\indent In an ongoing process a walker
following the iteration rules may get stuck close to the vicinity of one of
the unstable fixed points of the map. It has been shown both analytically
and numerically, that the PDF of escape times
from the vicinity of the fixed points decays as a power law \cite{Geisel1}.
To see this, one considers the dynamics in a half unit cell, say $0<x<1/2.$
Assume that at time $t=0$ the particle is on $x^{\ast }$ residing in the
vicinity of the fixed point $x=0.$ Close to the fixed point we may
approximate the map Eq. (\ref{n1}) with the differential equation $%
dx/dt=F\left( x\right) =ax^{z}.$ This equation is reminiscent of the equation
defining the $q$ generalized Lyaponov exponent \cite{Ananos}. The solution
is written in terms of the \textit{q exponential} function, $\exp _{q}\left(
y\right) \equiv \left[ 1+\left( 1-q\right) y\right] ^{1/\left( 1-q\right) }$
where $q=z$ and%
\begin{equation}
x\left( t\right) =x^{\ast }\exp _{z}\left[ ax^{\ast \left( z-1\right) }t%
\right] .  \label{n4}
\end{equation}%
We invert Eq. (\ref{n4}) to obtain the escape time from $x^{\ast }$ to a
boundary on $b (x^{\ast }<b<1/2)$ which is $t\simeq \int\limits_{x^{\ast
}}^{b}\left[ F\left( x\right) \right] ^{-1}dx$, using Eq. (\ref{n2}) we
obtain $t\simeq (1/a)\left[(x^{\ast 1-z})/(z-1)-(b^{1-z})/(z-1)\right] $
a $\ln _{q}$ ( i.e., the inverse function of $\exp _{q}$ )
behaviour. The PDF of escape times $\psi \left( t\right) $ is related to the
unknown PDF of injection points $\eta \left( x^{\ast }\right) ,$ through the
chain rule $\psi \left( t\right) =\eta \left( x^{\ast }\right) \left\vert
dx^{\ast }/dt\right\vert .$ Expanding $\eta \left( x^{\ast }\right) $ around
the unstable fixed point $x^{\ast }=0$ one finds that for large escape times 
\begin{equation}
\psi \left( t\right) \sim \frac{A}{\left\vert\Gamma \left( -\alpha \right)\right\vert }%
t^{-1-\alpha },\mbox{ \ }\alpha =\left( z-1\right) ^{-1}\mbox{\ \ }
\label{n5}
\end{equation}%
where $A$ depends on the PDF of injection points, namely on how trajectories
are injected from one cell to the other. The parameter $A$ is sensitive to
the detailed shape of the map while the parameter $z$ depends only on the
behaviour of the map close to the unstable fixed points. When $z>2$ the average escape time diverges
\cite{Zaslavski1,Geisel1,Geisel2}, and in turn yields aging \cite{Barkai1}, non-stationarity \cite{Grigolini},
anomalous diffusion \cite{Klafter,Korabel} and as we show here WEB.
\section{Weak Ergodicity Breaking} 
The map is said to be ergodic if for almost any initial condition (excluding paths starting at fixed
points) the path spends equal amounts of time in each cell in the limit of long measurement time. Namely
$t^{1}/t=1/L $, where $t^{1} $ is the time spent in one of the lattice cells and $t $ is the total
measurement time. Another method to quantify the dynamics is to consider the fraction of
number of visits per cell $n^{1}/n $ where $n^{1} $ is the number of visits in one of the lattice cells and $n $
is the total number of visits (or total number of jumps between cells). 
For maps with finite average escape time
$\left<\tau\right>=\int\limits_{0}^{\infty } \tau \psi \left( \tau \right) d\tau $, 
i.e., $z>2 $, the total measurement time $t \sim n\left<\tau\right> $ and the occupation time of one cell $t^{1}\sim
n^{1}\left<\tau\right> $ hence $n^{1}/n=1/L $, the equalities above should be understood in statistical sense.
We call $n^{1}/n $ the visitation fraction, which plays an important role in WEB.\newline
\indent More generally the fraction of occupation times (visitation fraction) in $m\leq L $ cells is $t^{m}/t
\left(n^{m}/n\right) $, where $t^{m} \left(n^{m}\right) $ is the occupation time (number of visits) in $m $
cells respectively. For ergodic systems the PDF of the fraction of occupation time is
\begin{equation}
f\left( \frac{t^{m}}{t}\right) =\delta \left( \frac{t^{m}}{t}-\frac{m}{L}%
\right),    \label{erg}
\end{equation}
and the visitation fraction is
\begin{equation}
\frac{n^{m}}{n}=\frac{m}{L}   \label{vis}
\end{equation}
in statistical sense, both equalities are valid only in the limit of long measurement time
($t\rightarrow \infty $). Note that the specific choice of the $m $ cells is not important.
Maps exhibiting strong non-ergodicity do not obey Eqs. (\ref{erg},\ref{vis}).
Dynamics obeying Eq. (\ref{vis}) but not Eq. (\ref{erg}) are said to be weakly non-ergodic.
As we will show when $z>2 $ WEB is found.\newline
\indent We start our analysis by discussing numerical simulations of the map with system of size $L=9 $.
\begin{figure}[tbp]
\begin{center}
\epsfxsize=80mm \epsfbox{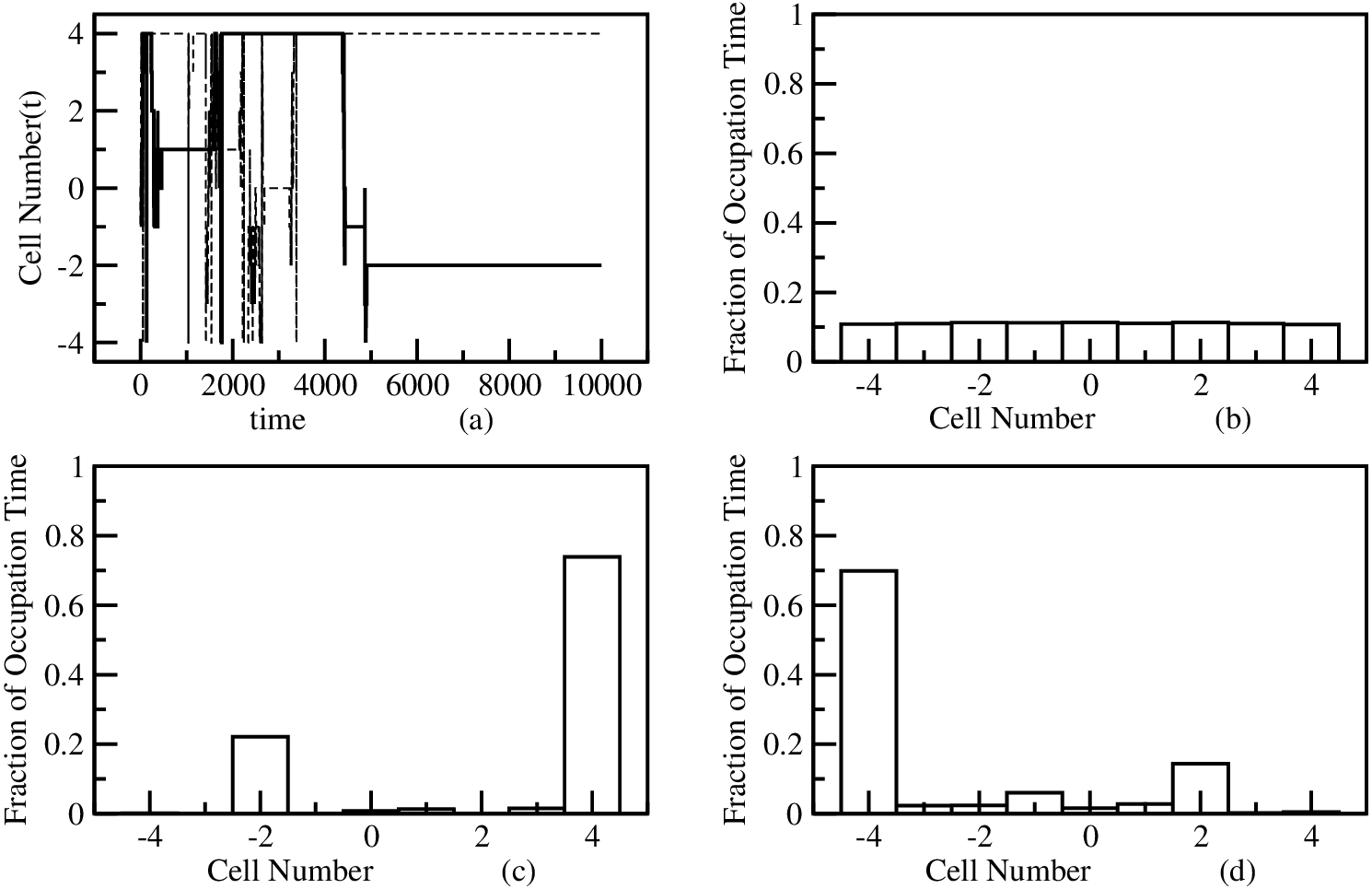}
\end{center}
\caption{ (a) Two paths (dashed and solid curves) generated according to the mapping rule
with z=3, for each path the particle gets stuck in one of the cells for a time which is
of the order of the measurement time. (b) Histogram of
the fraction of occupation for a path with z=1.5. For this ergodic case the fraction of occupation
times in the cells are equal. (c)+(d) The same as in (b), but this
time for the non-ergodic phase with z=3.}
\label{fig2}
\end{figure}
In Fig. \ref{fig2} (a) we show two paths generated according to the
mapping rule Eqs. (\ref{n2},\ref{n3}) with z=3. As one can see
each path gets stuck at one of the cells for a time which is of the order of the
measurement time. During the measurement the path visits all
the $L $ cells many times. In Fig. \ref%
{fig2} (b) we consider the case $z=1.5 $, we show that the fraction of
occupation time obeys Eq. (\ref{erg}) with $m=1 $ namely $t^{1}/t=1/L $, with $L=9 $. On the other hand in Fig. \ref{fig2} (c,d) we show
the fraction of occupation time histogram for two initial conditions obtained using $z=3 $. Since each path is localized in one of
the cells for a time which is of the order of the total measurement time, the histogram for this case looks
very different than that of the ergodic case and clearly Eq. (\ref{erg}) is not valid.
Histograms of the visitation fraction in each cell are presented in Fig. \ref{fig3},
for both ergodic ($z=1.2 $) (a)
and non-ergodic ($z=3 $) (b) maps. The Fig. demonstrates that for both cases Eq. (\ref{vis}) holds,
i.e., the visitation fraction in each cell is given by $1/L $. From Fig. \ref{fig2}
we see that the fraction of occupation time is random, hence we investigate its distribution.
We will find universal PDFs of occupation time, which generalize the ergodic rule Eq. (\ref{erg}).\newline
\begin{figure}[tbp]
\begin{center}
\epsfxsize=80mm \epsfbox{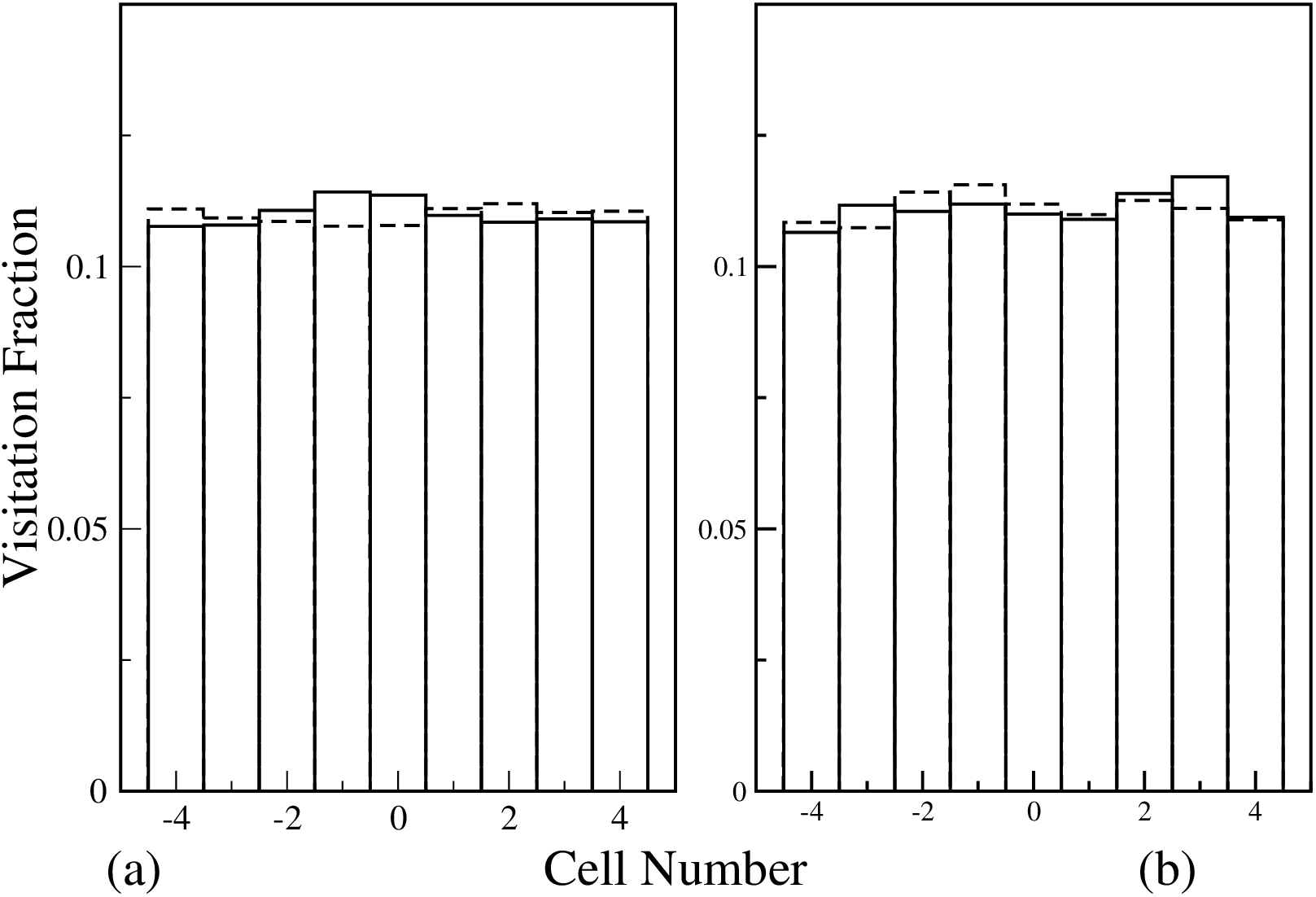}
\end{center}
\caption{ Histogram of the visitation fraction in each of the 9 lattice cells. 
(a) z=1.2 dashed and solid lines correspond to two initial conditions.
(b) The same as (a) but now z=3. In both
cases the visitation fractions in each cell are equal, with small fluctuations.}
\label{fig3}
\end{figure}
\indent The PDF for $t^{m}$( the occupation time of $m $ cells) in the case of diverging mean escape time is
now derived. In particular we show that the assumption Eq. (\ref{vis}) with power law waiting times, leads to nontrivial
PDFs of fraction of occupation times of the map. We denote by $f_{n,t}^{out}\left( t^{m}\right) $
the PDF\ of $t^{m},$ in the case that the path isn't within the $m $ cells at time $t $ and that during
the measurement time $t $ there were $n$ jumps.
Then
\begin{equation}
f_{n,t}^{out}\left( t^{m}\right) =\left\langle \delta \left(
t^{m}-\sum_{k=1}^{n_{i}}\tau _{k}^{m}\right) I\left( T_{n}<t<T_{n+1}\right)
\right\rangle ,  \label{C1}
\end{equation}%
where the summation in the delta function is over all sojourn times within the $m
$ cells, namely $\tau _{k}^{m} $ is the $k'th $
sojourn time within the $m $ cells, $T_{n}$ is the time after $n$ jumps and $%
I\left( x\right) $ is equal $1$ if $x$ is true and $0$ otherwise, the brackets denote average over all sojourn
times $\tau $. Double Laplace transform (LT) of Eq. (\ref{C1}) yields $\widetilde{f}_{n,s}^{out}\left( u\right)=$
$$ \left\langle 
\int\limits_{0}^{\infty }\int\limits_{0}^{\infty }e^{-ut^{m}}e^{-st}\delta
\left( t_{i}-\sum\limits_{k=1}^{n^{m}}\tau _{k}^{m}\right)I\left( T_{n}<t<T_{n+1}\right) dt_{i}dt%
\right\rangle $$
\begin{equation} 
=\frac{\widetilde{\psi }^{n^{m}}\left( u+s\right) \widetilde{\psi }%
^{n-n^{m}}\left( s\right) \left[ 1-\widetilde{\psi }\left( s\right) \right] 
}{s},
\label{C22}
\end{equation}
where $\widetilde{\psi }\left(s\right) $ is the LT of $\psi\left(t\right) $. We assumed that waiting times in
cells are not correlated. This and other assumptions are checked later using numerical simulations.
In a way similar to Eq. (\ref{C22}) we denote by $f_{n,t}^{in}\left( t^{m}\right) $ the PDF of $%
t^{m}$ in the case that the particle is within the $m $ cells at the end of the
measurement, then in double Laplace space we find
\begin{equation}
\widetilde{f}_{n,s}^{in}\left( u\right) =\widetilde{\psi }^{n^{m}}\left(
u+s\right) \widetilde{\psi }^{n-n^{m}}\left( s\right) \frac{1-\widetilde{%
\psi }^{n^{m}+1}\left( u+s\right) }{u+s}.
\label{C44}
\end{equation}
Considering an ensemble of initial conditions uniformly distributed in a cell,
the probability for the path to be within the $m $ cells at the end of the
measurement is $m/L,$ thus the double Laplace transformed PDF of the
occupation time of the cell, given that there were $n$ visits during the
measurement is
\begin{equation}
\widetilde{f}_{n,s}\left( u\right) =\frac{m}{L}\widetilde{f}_{n,s}^{in}\left(
u\right) +\left(1-\frac{m}{L}\right)\widetilde{f}_{n,s}^{out}\left( u\right).
\label{C5}
\end{equation}
Using Eqs. (\ref{C22},\ref{C44}) and the assumption $%
n^{m}/n=m/L$ Eq. (\ref{vis})we rewrite $\widetilde{f}_{n,s}\left( u\right) $ as
$$\widetilde{f}_{n,s}\left( u\right)=\left[ \frac{m}{L}\frac{1-\widetilde{%
\psi }\left( u+s\right) }{s+u}+\frac{L-m}{L}\frac{1-\widetilde{\psi }\left(
s\right) }{s}\right] $$
\begin{equation}
\left( \widetilde{\psi }^{m/L}\left( u+s\right) \widetilde{\psi }^{\frac{%
L-m}{L}}\left( s\right) \right) ^{n}.
\label{C66}
\end{equation}
Let $f_{t}\left(t^{m}\right) $ be the PDF of occupation time of the $m $ cells,
its double LT is given by
$$ \widetilde{f}_{s}\left( u\right)=\sum\limits_{n=0}^{\infty}\widetilde{f}_{n,s}\left(
u\right)$$
$$=\left[ \frac{m}{L}\frac{1-\widetilde{\psi }\left( u+s\right) }{s+u}+\frac{%
L-m}{L}\frac{1-\widetilde{\psi }\left( s\right) }{s}\right] $$
\begin{equation}
\frac{1}{1-\widetilde{\psi }^{m/L}\left( u+s\right) \widetilde{\psi }^{%
\frac{L-m}{L}}\left( s\right) }.
\label{C77}
\end{equation}
Taking the limit $u,s\rightarrow 0 $ (using the asymptotic behaviour of the escape time PDF $\widetilde{\psi }\left(
s\right) \sim 1-As^{\frac{1}{z-1} }$), and inverting the double LT one obtains
Lamperti PDF \cite{Lamperti} of fraction of occupation time
\begin{equation}
f\left(\frac{t^{m}}{t}\right)=\frac{\sin \left( \frac {\pi}{z-1}\right)}{%
\pi }\frac{R\left( \frac{t^{m}}{t}\right) ^{\frac{2-z}{z-1}}\left( 1-\frac{t^{m}%
}{t}\right) ^{\frac{2-z}{z-1}}}{R^{2}\left( 1-\frac{t^{m}}{t}\right) ^{\frac{2}{z-1}
}+\left( \frac{t^{m}}{t}\right) ^{\frac{2}{z-1} }+2R\left( 1-\frac{t^{m}}{t}%
\right) ^{\frac{1}{z-1} }\left( \frac{t^{m}}{t}\right) ^{\frac{1}{z-1} }\cos \left(\frac{\pi}{z-1}\right) } 
\label{delta}
\end{equation}
where $R=\frac{m}{L-m}$,
which is valid only when $t$ is large. Eq. (\ref{delta}) for $m=1$ is in agreement with the formalism developed in \cite{Bel}.
\begin{figure}[tbp]
\begin{center}
\epsfxsize=80mm \epsfbox{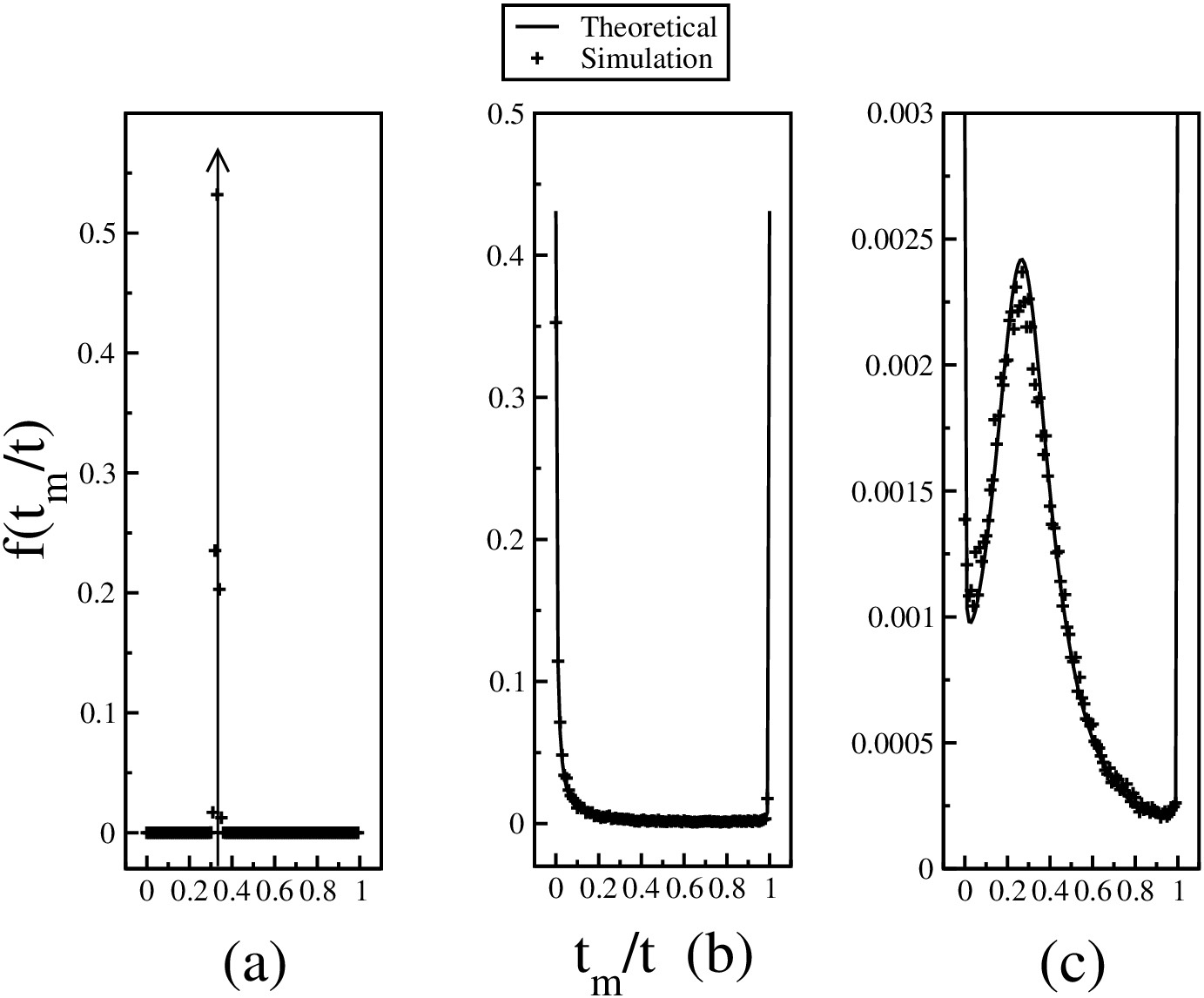}
\end{center}
\caption{ (a) The PDF of the fraction of
occupation time of m=3 cells out of L=9 cells for z=1.5, In this ergodic case the
PDF is very narrow and centered around m/L. (b) Same as (a), however here z=3. The PDF has a U shape,
with maxima at 0 and 1.
(c) Same as (a) with z=2.25. In this case a W shaped PDF is obtained, with a peak at the vicinity of m/L, but yet the maxima are at 0 and 1.
The curve is Eq. (\ref{delta}) without fitting. Only the parameter $z $ determines the shape of the PDFs, while detailed shape of the map is
unimportant. }
\label{fig4}
\end{figure}
Here we showed that the WEB condition
$n^{m}/n=m/L$ (i.e., the visitation fraction is equal to the probability of 
occupying the $m$ cells in \emph{ensemble} sense) and the power law behaviour of the sojourn time
PDF\ lead to the statistical law Eq. (\ref{delta}).\newline
\indent The assumptions made in the derivation of Eq. (\ref{delta}) are tested using numerical simulations of the
map. Plots of the PDF of fraction of occupation time for three different values of $z$ are shown in Fig. \ref{fig4}.
We used $10^{6}$ trajectories and measured the occupation time of $m=3 $ cells in a system of size $L=9 $. In the
ergodic case $z<2 $ the PDF is just a delta function centered at $m/L$ (see Fig. \ref{fig4} (a)). 
In the non-ergodic cases the maxima of $f\left(t^{m}/t\right)$ are at 0 and 1. These events
$t^{m}/t \approx 1 (t^{m}/t \approx 0)$ correspond to trajectories where the particle
occupies (does not occupy) one of the observed cells for the whole duration of the measurement, respectively.
In the non ergodic phase $z>2 $ two types of behaviors are found. 
For $z=2.25 $ the PDF of fraction of occupation time has a $W $ shape, with a peak in the vicinity of $m/L
$. While for $z=3 $, we find a $U $ shape PDF and the probability of obtaining fraction of 
occupation time equal to $m/L $ is almost zero. 
Comparison of theory Eq. (\ref{delta}) with the
numerical simulations of the map shows excellent agreement without any fitting, thus justifying our assumptions.
Note that if we limit our model to integer $z$ (i.e., assuming an analytical
behavior) then we either observe WEB ($z>2$) or the special border point between
WEB and ergodicity $z=2$. The latter point exhibits very slow convergence and
logarithmic corrections which should be the subject of future work.
\section{Conclusion} 
A detailed characterization of weak ergodicity breaking was established as follows:
(i) The phase space is not divided into mutually inaccessible regions, thus the system visits all
cells in its phase space for almost all (all but a set of measure zero) initial conditions.
(ii) The visitation fraction fluctuates very slightly in the limit $t\rightarrow \infty $ (note that $n$
and $n^{m}$ exhibit strong fluctuations).
(iii) The visitation fraction is determined by the probability of finding a member of an 
ensemble of particles within the given cell.
(iv) The total time the system spends in each cell fluctuates strongly, and is described by Eq. (\ref{delta}), implying that statistics of
occupation times depend only on the parameter $z $ describing the non-linearity and not on the shape of the map far from the fixed points
(e.g. $a $ in Eq. (\ref{n2})). A relation between weak chaos, deterministic
anomalous diffusion and weak ergodicity breaking was found. Since the former
behaviors are wide spread in dynamical systems, it is possible that the
non-ergodic dynamical scenario might emerge in other fractional dynamical
systems, and then the present concepts naturally lead to a new type of weakly
non-ergodic statistical mechanics.

\acknowledgments
EB thanks the Center for Complexity Science Jerusalem for support.

\end{document}